\documentclass[prl,twocolumn, aps,amssymb,footinbib,showpacs,superscriptaddress]{revtex4-1}
\pdfoutput=1

\usepackage{graphicx}
\usepackage{amsmath,amssymb}
\usepackage{cancel}
\usepackage{color}
\usepackage{bm}
\usepackage{verbatim}

\newcommand{\ket}[1]{\ensuremath{\left| #1 \right\rangle}}

\def \up {\uparrow}
\def \down {\downarrow}

\def \half {\tfrac{1}{2}}

\begin{document}

\title{Continuous Preparation of a Fractional Chern Insulator}

\author{M. Barkeshli}
\affiliation{Microsoft Station Q, Elings Hall, University of
  California, Santa Barbara, CA, 93106, USA}

\author{N. Y. Yao}
\affiliation{$^{1}$Department of Physics, University of California Berkeley,  Berkeley, CA 94720, U.S.A.}

\author{C. R. Laumann}
\affiliation{Department of Physics, University of Washington, Seattle, WA, 98195, USA}

\begin{abstract}
We present evidence of a direct, continuous quantum phase transition between a Bose superfluid and the $\nu=1/2$ fractional Chern insulator in a microscopic lattice model.
In the process, we develop a detailed field theoretic description of this transition in terms of the low energy vortex dynamics. 
The theory explicitly accounts for the structure of lattice symmetries and predicts a Landau forbidden transition that is protected by inversion. 
That the transition is continuous enables the   quasi-adiabatic preparation of the fractional Chern insulator in non-equilibrium, quantum optical systems. 

% An exciting prospect in the study of synthetic quantum optical systems is the possibility of realizing 
% exotic topological phases of matter, such as the Laughlin fractional quantum Hall (FQH) states, and their
% lattice analogs, the fractional Chern insulators (FCI). 
% A major technical challenge is the question of preparing such strongly-correlated ground states. 
% One approach to this is to start with a weakly correlated state, such as a Bose superfluid, and 
% tune through a continuous quantum phase transition to the Laughlin state. 
% In this paper, we present two main advances. We present a microscopic 
% model, relevant to a two-dimensional lattice of nitrogen-vacency defects in diamond, which 
% exhibits a novel qualitative signature predicted by the field theory of the continuous transition 
% between the superfluid and the 1/2-Laughlin state. Furthermore, we develop a novel field theory of the 
% transition that takes into account the role of lattice symmetries and half-filling of the lattice.
\end{abstract}
%\pacs{TODO 73.43.Cd, 05.30.Jp, 37.10.Jk, 71.10.Fd}
%\keywords{TODO ultracold atoms, polar molecules, gauge fields, flat bands, superfluid, supersolid, dipolar interactions}
\maketitle

% TODO Norm's figures and supp on numerics/microscopics
The canonical examples of topological order are provided by the fractional quantum Hall states, conventionally found in two-dimensional electron gases \cite{tsui1982,laughlin1983}. 
Their lattice cousins, the fractional Chern insulators (FCI), naturally arise when strongly interacting particles
inhabit flat, topological band-structures
% \cite{Roy:2011bz} This is the viewpoint
\cite{Kalmeyer:1987id,Parameswaran:2012cu,Wang:2011cy,Liu:2012ek,Sheng:2011iv,Regnault:2011bu,McGreevy:2012ek,Sun:2011dk,Tang:2011by,Neupert:2011db,Yao:2012fn,Moller:2012ej}.
Effective microscopic Hamiltonians whose ground states realize such phases have been numerically identified in synthetic quantum systems, ranging from ultracold gases in optical lattices to ensembles of solid-state defects \cite{Yao:2013eg,Cooper:2013jg,Bennett:2014}. 
On the experimental front, \cite{Aidelsburger14}  have recently loaded $^{87}$Rb into the topological, nearly-flat band of a Hofstadter model.

%Quantum optical platforms offer novel ways to probe the underlying topological structures, for example by directly
%imaging fractionalized excitations in real space. They provide access to topological phases distinct from those
%conventionally realized in electronic systems, such as the bosonic Laughlin states. The lattice-scale densities involved
%also give rise to rich physics due to the interplay between lattice symmetries and 
%topological order.
%
Unlike typical condensed matter systems, quantum optical proposals of topological phases represent driven, non-equilibrium implementations 
in an effective Hamiltonian picture. 
Thus, even if an appropriate Hamiltonian can be realized, guiding the system to its groundstate is still a major challenge. 
Often, one cannot simply ``cool'' by decreasing the temperature of a surrounding bath. 
One approach to this problem is provided by quasi-adiabatic preparation, wherein the correlated ground state is reached from a simple initial state by slowly tuning the Hamiltonian parameters. 
In the case of FCIs, natural starting states include superfluids (SF) and charge-density wave (CDW) insulators, as these often arise in close proximity to the FCI state of interest \cite{Yao:2013eg}.

Quasi-adiabatic preparation requires that any quantum phase transition between the initial and final state be continuous. 
A system tuned through a first order transition would need to be ramped exponentially slowly in system size 
to avoid being stuck in a metastable high energy state \cite{Laumann:2012hu,Schutzhold:2006gt}.
On the other hand, continuous quantum phase transitions allow for two possibilities:  1) strictly adiabatic preparation with ramp time scaling as a power law in system size \cite{Cardy:1984p10764,Sondhi:1997p8125,Sachdev:2011} or 2) quasi-adiabatic preparation with a final state energy density scaling  as an inverse power law with the ramp time \cite{Kibble:1976fm,Zurek:1985ko,Chandran:2012fh}.
Unfortunately, there is relatively little known regarding quantum phase transitions between conventional and fractional phases as such transitions lie beyond the Ginzburg-Landau paradigm \cite{wen04}.

Field theories of possible critical points between Laughlin fractional quantum Hall states and Mott insulators were studied in \cite{zhang1989,wen1993,chen1993}.
Meanwhile, a theory of a superfluid to bosonic $\nu=1/2$ Laughlin state was recently constructed in \cite{barkeshli2012sf}. 
%Continuous transitions Early field theoretic work found continuous transitions between the Laughlin fractional quantum Hall states and Mott insulators %\cite{ZHK,WenWu,MPA}. More recently, a theory of the transition between the superfluid and the $\nu=1/2$ bosonic Laughlin state was constructed %\cite{MBMcGreevy}, wherein the direct transition between these two states is protected by some point group symmetry, such as inversion. In the absence of inversion, %the direct transition splits into two transitions with an intervening Mott insulator. 
All of these theories assume that any additional lattice symmetries are preserved throughout the phase diagram. They  require the bosons to be at integer filling and cannot describe CDW order. Moreover, to date, none of these continuous transitions has been established in any microscopic model, as second order phase transitions are difficult to characterize in the small systems amenable to numerical study.

\begin{figure}
\centering
\includegraphics[width=3.40in]{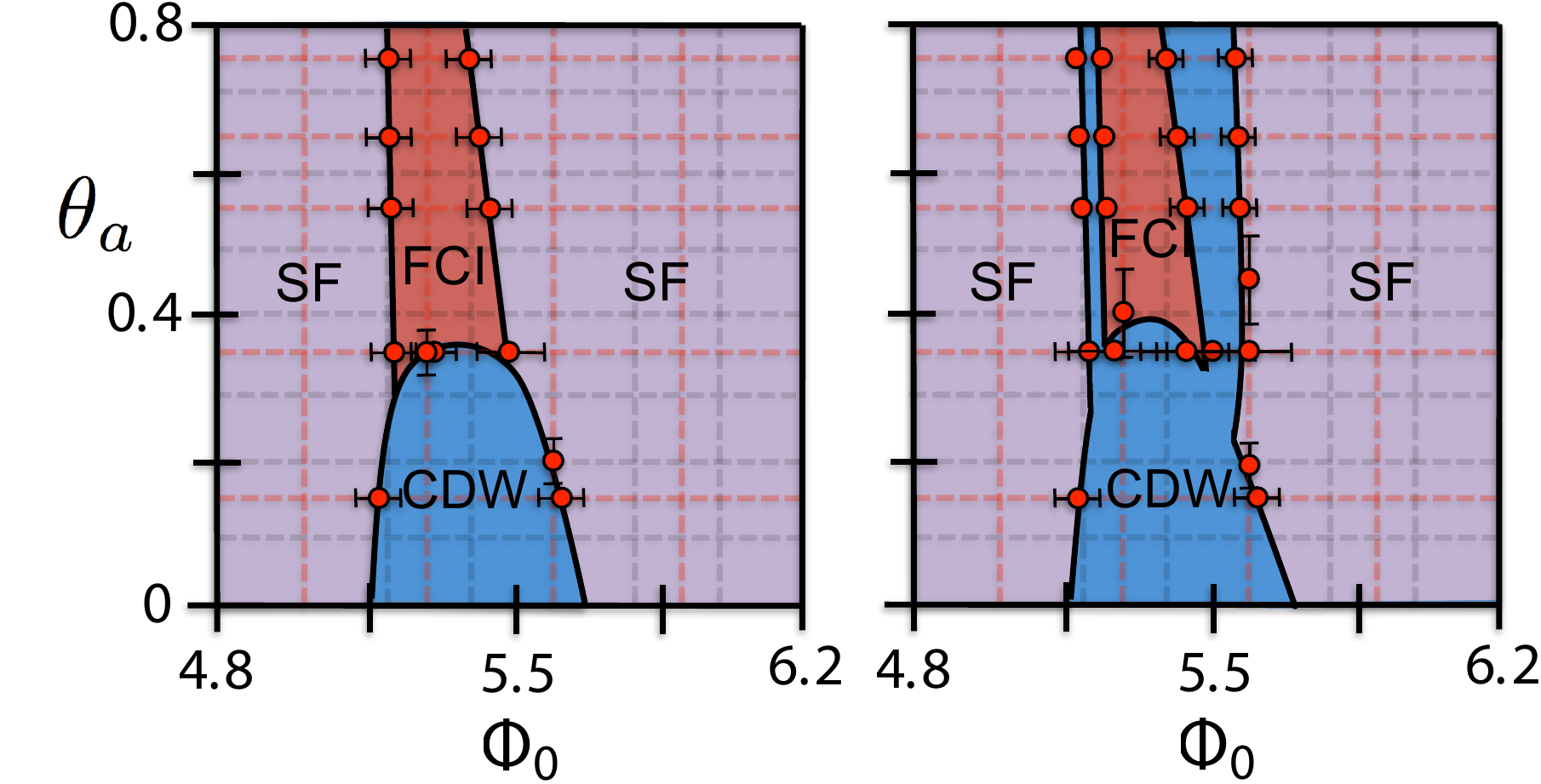}
\caption{ (a) Two parameter phase diagram of the driven NV model as determined by exact diagonalization of Eq.~\eqref{eq:ham_hcboson}. (b) Phase diagram in the presence of microscopic inversion symmetry breaking parameter $g = 0.2$. The $(\pi,\pi)$ CDW insulator extends in two fingers which split the SF $\leftrightarrow$ FCI transition, showing that the underlying transition at $g=0$ is continuous and protected by inversion symmetry. 
Spectra and structure factors collected on coarse grey grid sites; full diagnostics (see text) calculated on 1-D (red) cuts at spacing of $0.01$. 
Markers with errorbars indicate regions where diagnostics were ambiguous. 
Markers without errorbars indicate ambiguous regions narrower than marker size.
 % \textcolor{red}{grey lines are at 0.1 in theta-a and 0.2 in phi0. for each such point in the grey subgrid, we calculated structure factor and spectrum. along the red lines we go in steps of 0.01 in both directions and at each point we also calculate the SF response for twists along x and y. In the y direction, the twist is 4pi (in 20 steps) and in the y-direction the twist is 2pi in 10 steps. At each step, we also calculate the structure factor and spectrum. For these red lines we calculate sigma-xy as well.}
 } 
\end{figure}

In this Letter, we report two main advances. First, we establish the presence of a direct continous 
transition between a superfluid and a $\nu=1/2$ FCI state in a microscopic model of interacting spins.  
We do this by showing that the direct superfluid - FCI transition splits into two transitions when we perturbatively break inversion symmetry.
%in the microscopic model, as predicted by the effective field theory of the continuous transition. 
Since first order phase transitions are insensitive to perturbations, the splitting of the transition implies that it must be continuous. This qualitative
signature 
%observing qualitative signatures associated with breaking inversion symmetry, as predicted by field theoretic considerations. 
avoids the usual difficulty associated with finite-size scaling  in small systems. 
Second, we develop a detailed field theoretic description of this transition in terms of the low-energy vortex fields. 
This description naturally accommodates the spontaneous breaking of lattice symmetry in the Mott-insulating CDW state at half-filling.

\begin{figure}
\centering
\includegraphics[width=1.0\columnwidth]{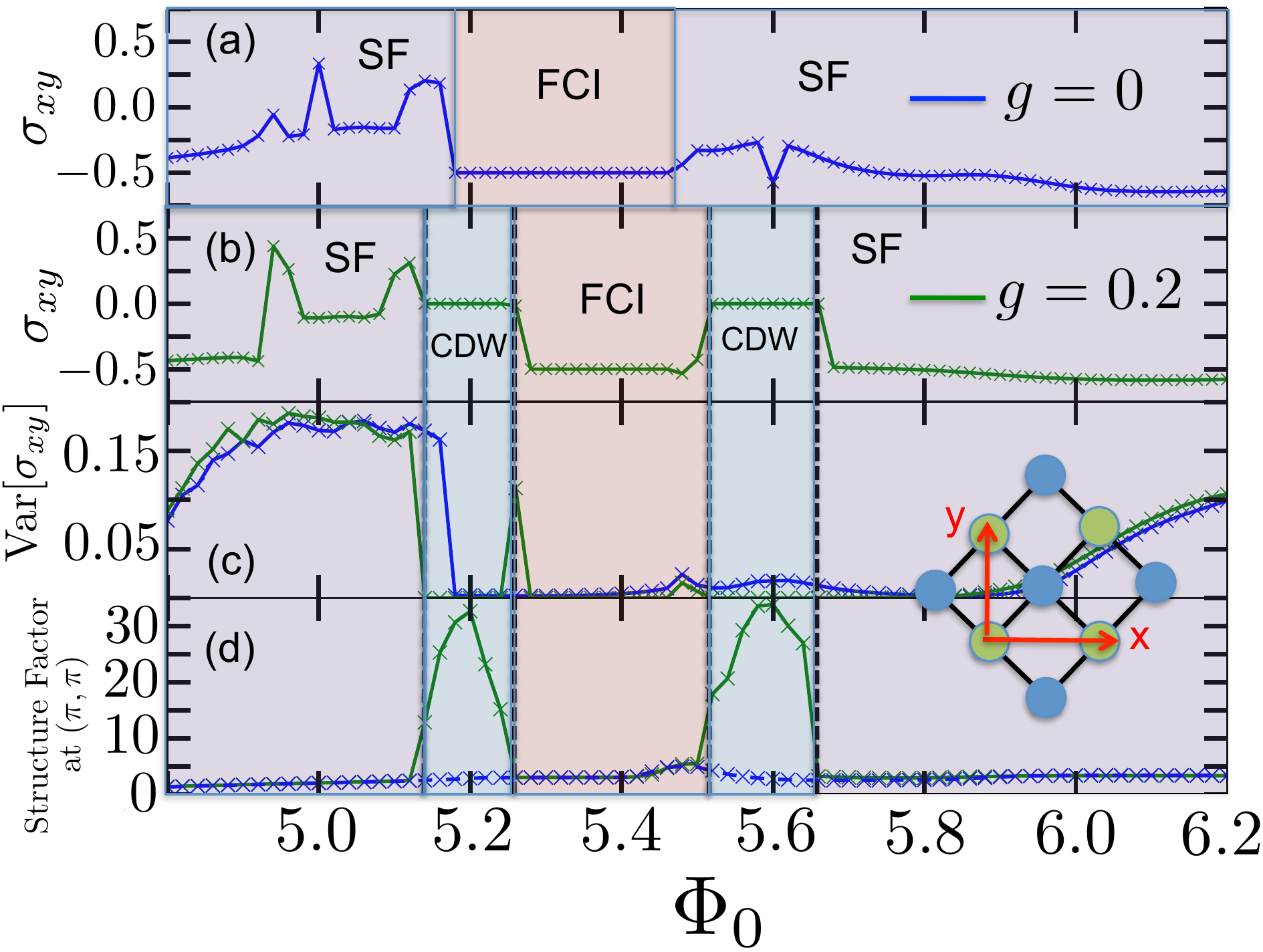}
\caption{Numerical diagnostics on a cut of the phase diagram at $\theta_a = 0.75$ calculated at $N_{sites} =32$, $N_{particles} = 8$. (a)  Berry curvature $\sigma_{xy}$ averaged over the boundary condition torus for $g = 0$. In the SF, $\sigma_{xy}$ is not quantized, while in the FCI, it is precisely $-0.5$. (b) Analogous with inversion broken $g=0.2$. The intervening CDW exhibits $\sigma_{xy}=0$. (c) Fluctuations of the Berry curvature sampled on a $10 \times 10$ grid in the boundary condition torus. Notice that fluctuations are heavily suppressed in the insulating phases while the gapless-ness of the SF causes a large variance. (d) Real space structure factor at $k = (\pi, \pi)$. Both the SF and FCI are translation invariant while the CDW exhibits strong ordering. (inset) Depicts the two-site unit cell square lattice and its primitive vectors. 
} 
\end{figure}

\emph{Microscopic Model}---We study the microscopic phase diagram of a two-dimensional  square lattice of Nitrogen-Vacancy (NV) defects in diamond. Our model is closely related to previous proposals for realizing FCI states in ultracold polar molecules \cite{Yao:2013eg}. We will briefly sketch the main ingredients below (for details see supplementary information \cite{suppinfo}).
Each NV constitutes a spin one ($S=1$) electronic degree of freedom and interactions occur via the magnetic dipole-dipole interaction,
\begin{equation}
H_{dd} = \frac{1}{2} \sum_{i\neq j}  \frac{\kappa}{R_{ij}^3}   \left [  {\bf S}_i  \cdot {\bf S}_j - 3({\bf S}_i \cdot {\bf \hat{R}}_{ij})({\bf S}_j \cdot {\bf \hat{R}}_{ij}) \right ],
\end{equation} 
where, $\kappa = \mu_0/(4\pi)$ and ${\bf R}_{ij}$ connects sites $i$ and $j$  \cite{Childress:2006km}. 
%%
%
%
%
%
%and is strongly coupled, via hyperfine interactions, to an $I=1/2$ nuclear spin ($^{15}$N) \cite{Childress:2006km}. Interactions between NVs occur   via the magnetic dipole-dipole interaction,

Taking into account the  zero-field splitting, an applied magnetic field, the hyperfine interaction, and electromagnetic radiation (optical dressing \cite{Yao:2013eg}), one finds that the system is described by an effective Hamiltonian,
\begin{equation}
	\label{eq:ham_hcboson}
H_B =  -\sum_{ij} t_{ij} d_i^{\dagger} d_j + \frac{1}{2}\sum_{i \neq j} V_{ij} n_i n_j.
\end{equation}
Here, $d_i^{\dagger}$ are conserved hardcore bosons which emerge as dark states of the optical dressing. Both the dynamics and interactions of these bosons are governed by the dipolar interaction. Thus, $ t_{ij}$, $V_{ij} $ are matrix elements of $H_{dd}$ in the dark state subspace;  they exhibit $1/R^3$ tails and strong anisotropy  \cite{Bennett:2014,suppinfo}.
In addition to boson number conservation, $H_B$ is symmetric under lattice translations and spatial inversion. We note that the elliptical polarization of the electromagnetic radiation directly breaks time-reversal symmetry \cite{Yao:2013eg,suppinfo}.

An FCI can be realized in this system with two main kinetic ingredients: the single boson bands ought to be ``flat'', such that their dispersion is small relative to the interactions, and they ought to carry a non-trivial Chern number. 
Such topological flat bands may be achieved by using different optical dressing parameters on the $a$ and $b$ sites of a two-site unit cell (green and blue, inset Fig.~2); this amounts to defining the hardcore boson slightly differently on the $a$ and $b$ sublattices \cite{Yao:2012fn}.

We now consider the many-body phases which arise at filling fraction $\nu =1/2$ per unit cell (\emph{i.e.} $1/4$ particle per site) in a topological flat band regime. 
The phase diagram depicted in Fig.~1a is calculated using exact diagonalization for sizes up to $N_{sites} = 36, N_{particles} = 9$. 
Two microscopic parameters are varied: $\Phi_0$  is the azimuthal angle of the NV axis relative to the lattice plane and $\theta_a$ is a microscopic dressing parameter.
Roughly speaking,  $\theta_a$ controls the magnitude of the effective  interaction $V_{ij}$ (with $\theta_a \rightarrow 0$ giving the strongest interactions), while $\Phi_0$ controls the amount of  band dispersion. 
These qualitative differences in the microscopics yield a rich phase diagram exhibiting both conventional and topological phases (Fig.~1a). 

A $\nu=1/2$ bosonic Laughlin FCI arises where the dispersion is flattest and the dipolar tail of the interaction is weak. 
Turning up the interactions by varying $\theta_a$ causes the system to spontaneously break the lattice translational symmetry and form a commensurate CDW insulator at momentum $(\pi,\pi)$. 
Tuning away from the flat band regime by adjusting $\Phi_0$ leads to a phase transition into a superfluid, consistent with the microscopics being dominated by band dispersion. 
We identify these phases numerically with five diagnostics: i) ground-state degeneracy, ii) spectral flow under magnetic flux insertion (superfluid response), iii) real-space structure factor $\langle n(R) n(0) \rangle$, iv) the many-body Berry curvature $\sigma_{xy}=  \frac{1}{2\pi} \int \int F(\theta_x, \theta_y) d\theta_x d\theta_y $ with $F(\theta_x, \theta_y) = \text{Im} ( \langle \frac{ \partial \Psi}{\partial \theta_y} |  \frac{ \partial \Psi}{\partial \theta_x} \rangle - \langle \frac{ \partial \Psi}{\partial \theta_x} |  \frac{ \partial \Psi}{\partial \theta_y} \rangle)$
\footnote{We use the symbol $\sigma_{xy}$ for the Berry curvature even though it only agrees with the Hall conductance in the gapped phases}, and v) (for the FCI), Laughlin quasi-hole counting \cite{Regnault:2011bu, suppinfo}. 

The above diagnostics unambiguously determine the phases deep within each phase. 
The phase boundaries sketched in Fig.~1a correspond to the regions where the diagnostics become ambiguous due to the finite size crossovers. 
%These also correspond to the regions where the ground state energy varies most sharply as a function of the control parameters $\theta_a$ and $\Phi_0$, see Fig.~2a for a typical example.
The error bars in the phase diagram indicate the width of the crossover region as observed in the five diagnostics.

Whether the transition is continuous or first order is hard to extract directly by conventional methods from such small size numerics.
So we use a trick: 
the known critical theories describing the direct SF$\leftrightarrow$FCI transition require a discrete symmetry, such as inversion, to protect them. 
Thus, if breaking inversion perturbatively in the microscopic model introduces a Mott insulator between the SF and FCI phases we can conclude that the underlying transition was continuous.

To test this, we introduce a weak staggering $g$ to the horizontal nearest neighbor hopping, $t_{i,i+\hat{x}} \to (1+g)^{s_i} t_{i,i+\hat{x}}$, where $s_i$ is $0$ ($1$)  on the $a$ ($b$) sublattice.  
We have investigated the phase diagram with $g=0.2,0.3,0.4$; the phase diagram with $g=0.2$ is shown in Fig.~1b using the same numerical diagnostics as before (Fig.~2) \cite{suppinfo}. 
The introduction of staggering indeed splits the FCI to SF transition revealing an intermediate CDW insulator.
We view this as strong evidence that the transition at $g = 0$ is continuous and described by the field theory we develop below. 

\begin{figure}
		\includegraphics[width=3.3in]{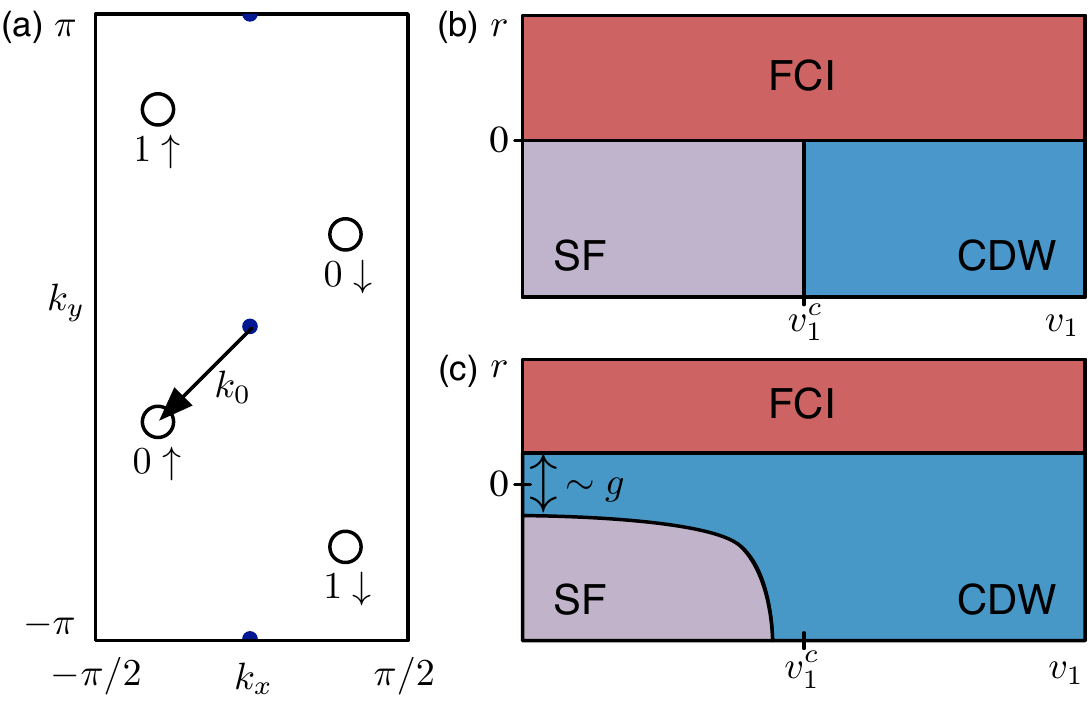}
	\caption{
	(a) Magnetic Brillouin zone for vortex fields $\phi^v_{l\alpha}$ in Landau gauge. Circles indicate dispersive minima and where the slow vortex fields are defined.
	(b) Two parameter phase diagram of theory \eqref{dual2sfcdwfqh} without inversion breaking. Slice in $r,v_1$ holding 
%$v_2<0, v_3>0,w_1>0, w_2<0, w_3>0$ 
$v_2 < v_3 < 0$, $w_2 < 0$ and $w_1, v_3, w_3 > 0$
and $u>0$ large enough to stabilize the potential, yields an inversion breaking CDW with $(\pi,\pi)$ ordering and a superfluid with $(\pi,\pi)$ current order.
	(c) Same phase diagram with $g \ne 0$ breaking inversion.
}
	\label{fig:gl-fqhcdw}
\end{figure}

%%%%%%%%%%%%%%%%%%%%%%%%%%%%%%%%%%%%%%%%%%

\emph{Field theory---}In order to capture the phase transitions seen in Fig.~1, any long-wavelength description must be able to simultaneously accommodate a $\nu=1/2$ Laughlin state, the superfluid and the spontaneous breaking of lattice symmetry in the CDW insulator. 
Previous work \cite{barkeshli2012sf} considered the case where the Mott insulator is at integer filling and thus need not break  translational symmetry.
Here, we will present an alternate theory for  bosons at half-integer filling, 
which takes into account the fact that the CDW insulator must \emph{spontaneously} break lattice symmetry \cite{Lieb:1961dn,Hastings:2004cd}.
\emph{En passant,} our new theory provides a physical representation of the transition which emphasizes the role of vortex dynamics. 

We begin by briefly reviewing the effect of half-filling on the vortices of a superfluid state on a rectangular lattice \cite{balents2005,Lannert:2001jk}.
The vortices see the original particles as magnetic flux 
quanta \cite{dasgupta1981,fisher1989} and thus, on average, feel half a flux quantum per plaquette (of the dual lattice). 
This requires the translational symmetries of the vortex theory to be augmented by a gauge transformation.
The resulting $T_x$ and $T_y$ operators satisfy the ``magnetic'' translation algebra $T_x T_y = - T_y T_x$. 
The vortex bandstructure must have an even number of minima, protected by this translation algebra. 
If these minima are not at inversion symmetric points in the magnetic Brillouin zone, then inversion symmetry
$\mathcal{I}$ requires that the number of minima be a multiple of four (Fig.~3a).

In the minimal case there are four such minima at momenta $\pm k_0, \pm k_0 + (0,\pi)$ (in Landau gauge). 
A soft-mode expansion of the vortex field near these minima leads to four flavors of vortices which 
we label $\phi^v_{l\alpha}$ for $l=0,1$ and $\alpha = \uparrow,\downarrow$, as in Fig.~3a.
The symmetry operators act as follows:
\begin{align}
        \label{eq:symphiv}
\mathcal{I}&: \phi^v \rightarrow \tau_x \phi^v 
\nonumber\\
T_x&: \phi^v \rightarrow e^{ik_0\cdot\hat{x} \tau^z} \sigma^x \phi^v 
\nonumber\\
T_y&: \phi^v \rightarrow e^{ik_0\cdot\hat{y} \tau^z} \sigma^z \phi^v
\end{align}
where the $\tau$ ($\sigma$) Pauli matrices act on the $\alpha$ ($l$) index and $k_0$ is the momentum of the $0\up$ field.

In the superfluid state, all of these vortices are uncondensed. 
When any combination of them condenses, the superfluid order is
destroyed and the translation symmetry is broken, leading to insulating density wave states \cite{balents2005,suppinfo}. 
Remarkably, the $\nu=1/2$ Laughlin  state arises  when  the vortices form an integer quantum Hall state \cite{barkeshli2013csl,senthil2013,lu2014quantum,grover2013quantum,barkeshliHLR}.
This motivates the following field theory which can interpolate between the FQH, superfluid, and CDW states:
\begin{align}
\label{dual2sfcdwfqh}
\mathcal{L} &= \frac{1}{2\pi} A_e \partial a + \frac{1}{2\pi} b^\up \partial b^\down
- \frac{1}{2\pi} a \partial (b^\up + b^\down) 
\nonumber \\
&+ \sum_l |(\partial - i b \tau_x) \phi_{l}|^2 - V(\{\phi_l\}),
\end{align}
where the notation $a \partial b \equiv \epsilon^{\mu\nu\lambda} a_\mu \partial_\nu b_\lambda$.
Here, $a$ and $b^\alpha$ are internal $U(1)$ gauge fields minimally coupled to the  complex scalar fields $\phi_{l\alpha}$; $
A_e$ represents a background external gauge field used to probe the underlying boson current $j^\mu = 1/2\pi \epsilon^{\mu\nu\lambda}\partial_\nu a_\lambda$.
The Chern-Simons terms bind a flux quantum of $b^{\up/\down}$ to $\phi_{l\down/\up}$. These flux-$\phi_{l\alpha}$ composites represent the original vortex fields $\phi_{l\alpha}^v$.
Under the action of the lattice symmetries, $\phi_l$ can be taken to transform as $\phi_l^v$ in Eq.~\eqref{eq:symphiv}, while the gauge fields $b$ are invariant under $T_x, T_y$ and swap under $\mathcal{I}$.

The potential term $V = r \phi^\dagger \phi +  V_4 + \cdots$ includes all other terms 
compatible with the physical and gauge symmetries. 
At quartic order, there are seven couplings,
%\begin{widetext} r \sum_l  |\phi_l|^2 + 
\begin{align}
        \label{glpot}
V_4 &= u (\phi^\dagger \phi)^2 + 
 v_1 \sum_l |\phi_{l\up} \phi_{l\down}|^2 
+ v_2  \sum_\alpha |\phi_{0\alpha} \phi_{1\alpha}|^2 \nonumber \\
&+ v_3 (|\phi_{0\down} \phi_{1\up}|^2 + |\phi_{1\down}\phi_{0\up}|^2) 
+w_1 \sum_\alpha \phi_{0\alpha}^{*2}\phi_{1\alpha} \nonumber \\ %+ \textrm{c.c.} )+ 
&+ w_2 \phi_{0\up}^* \phi_{1\up} \phi_{0\down}^* \phi_{1\down} 
+ w_3 \phi_{0\up}^* \phi_{1\up} \phi_{1\down}^* \phi_{0\down} %+ \textrm{c.c.} )
+ \textrm{c.c.}
\end{align}
%\end{widetext}
This theory Eqs.~(\ref{dual2sfcdwfqh},\ref{glpot}) is one of the central results of the Letter. 
It is capable of describing all three phases found in the 
microscopic model: 
(1) When  $\phi_{l\alpha}$ are uncondensed ($\langle \phi_{l \alpha} \rangle = 0$)
they can be integrated out, yielding the effective theory of the $\nu=1/2$ 
Laughlin state \cite{wen04}.
(2) If one of the $\phi_{l\alpha}$ condenses, $b^\alpha$ is gapped by the Anderson-Higgs mechanism; 
the resulting theory describes a Mott insulator which, as shown below, breaks translation symmetry.
(3) If both $b^\alpha$ gauge fields are Higgsed, the resulting theory $\mathcal{L} = 1/(2\pi) A_e \partial a + (\partial a)^2 + \cdots$ is the usual dual description of a superfluid. 

The pattern of inversion and translation symmetry breaking in these phases follows from the behavior of the simplest gauge-invariant bilinears in the $\phi$ fields:
\begin{align}
\mathcal{O}_{0,0}^{\alpha} &\equiv \phi_\alpha^\dagger \phi_\alpha
\;\;\;
\mathcal{O}_{\pi,0}^\alpha \equiv \phi_\alpha^\dagger \sigma^z \phi_\alpha
\nonumber \\
\mathcal{O}_{0,\pi}^\alpha &\equiv \phi_\alpha^\dagger \sigma^x \phi_\alpha
\;\;\;
\mathcal{O}_{\pi,\pi}^\alpha \equiv \phi_\alpha^\dagger \sigma^y \phi_\alpha.
\end{align}
The operators $\mathcal{O}_{k_x,k_y}^\alpha$ carry momentum $(k_x,k_y)$. 
The linear combination $\mathcal{O}_{k_x,k_y}^\pm \equiv \mathcal{O}_{kx,ky}^\up \pm \mathcal{O}_{k_x,k_y}^\down$ is inversion even (odd).
Depending on which $\mathcal{O}_{k_x,k_y}^\pm$ acquire expectation values, we can determine how translation and inversion 
are broken 
\footnote{
We note that the original vortex fields $\phi^v$ carry incommensurate momenta at $k_0$, so their condensation could produce incommensurate ordering at momenta $2k_0$. 
Identifying such incommensurate orders here requires knowledge of the correlators of the monopole operators $M_\alpha$, a task beyond the scope of this paper.
}.

Figure 3b shows a particular 2-parameter slice of the mean-field phase diagram of Eq.~\eqref{dual2sfcdwfqh} which shows direct continuous transitions between the FCI $\leftrightarrow$ SF and FCI $\leftrightarrow$ CDW phases, along with a continuous triple point terminating the first order line separating the SF $\leftrightarrow$ CDW phases.
The CDW order is at momentum $(\pi, \pi)$, as seen in the numerics, while the superfluid has $(\pi, \pi)$ current order. 
% Comment this is a funny first order line -- it seems not to extend in the inversion breaking direction where it is really a second order sheet. I think this 
%
The leading inversion breaking potential, $V = g \phi^\dagger \tau^z \phi$, splits the direct FCI $\leftrightarrow$ SF transition 
by an intervening CDW  with width proportional to $g$ as in Fig.~3c.
The topology of these phase diagrams matches that observed numerically in Fig.~1.

Similar phase diagrams arise in other regions of the coupling space; in all cases, the insulators exhibits commensurate density order and the SF breaks a lattice symmetry.
Likewise, a superfluid living in a band structure with non-inversion symmetric minima will either condense into a standing wave or break inversion
The microscopic dispersion from Eq.~\eqref{eq:ham_hcboson}  indeed exhibits non-inversion symmetric minima, but the small accessible system sizes prevent us from  verifying the symmetry breaking pattern in the SF.

%It is possible to develop an alternative theory of the triple point, such that all transitions, including the $SF \leftrightarrow CDW$ transition, are continuous at the mean-field level. In such a theory, the inversion related vortex fields merge at the triple point and acquire an anisotropic dispersion. We leave the development of such a multicritical theory for future work. 

In summary, we have constructed a critical field theory that  describes transitions between FCI $\leftrightarrow$ CDW  $\leftrightarrow$ SF, accommodating both  spontaneous symmetry breaking and topological order.
 Surprisingly, this theory is realized in a microscopic model of coupled electronic and nuclear spins as arise in an engineered lattice of NV defects.
 While our microscopic study has focused on NVs, the universal physics predicted by the field theory should be applicable to phase transitions in ultracold atomic systems \cite{Aidelsburger14},  polar molecules \cite{Yao:2012fn,Yao:2013eg} and Rydberg ensembles \cite{peter2014topological}. 
 In such systems, we predict that the quasi-adiabatic preparation of a fractional state can occur with energy density
 \begin{align}
\epsilon \sim \tau^{-\frac{3\nu}{\nu+1}}
\end{align}
where $\tau$ is the ramp time and $\nu$ is the correlation length exponent of the field theory  \cite{kibble1976topology,zurek1985cosmological,chandran2012kibble}. We leave the precise calculation of $\nu$ to future work, but note that in the absence of gauge fluctuations, $\nu \approx 0.7$ \cite{chaikin2000principles} as for a two-component XY transition. 
For small finite size systems, we also expect the gap to close as $\sim 1/L$  since the dynamical critical exponent is $z=1$. 
This opens the door to preparing fractionalized states in near term quantum optical simulators. 

\paragraph{Acknowledgements---} We would like to acknowledge stimulating discussions with A. Vishwanath, A. Chandran, S.D. Bennett, A.V. Gorshkov, M.D. Lukin, J. McGreevy, S.A. Parameswaran, T. Senthil and S.L. Sondhi. CRL acknowledges the hospitality of the Perimeter Institute.

\bibliography{TIFQHCDW}

\appendix

\section{Microscopics and Numerical Diagnostics}

Here, we provide a  description of the microscopic spin model underlying the numerics presented in the maintext. To be specific, we consider Nitrogen-Vacancy defect centers in diamond. The electronic ground state of each NV center 
is a spin-1 triplet described by the Hamiltonian,
\begin{equation}
\label{eq:Hnv}
	H_{NV} = D_0 S_z^2 + \mu_e B S_z,
\end{equation}
where $D_0 = 2.87$ GHz is the zero
field splitting,
$\mu_e = - 2.8$ MHz/Gauss is the
electron spin gyromagnetic ratio, and
$B$ is a magnetic field applied parallel
to the NV axis. This electronic spin is coupled via hyperfine interactions to the $I=1/2$ nuclear spin of the $^{15}$N impurity via
\begin{equation}
\label{eq:hyperfine}
	H_{HF} =  A_{\parallel} S_z I_z + A_{\perp} (S_x I_x + S_y I_y),
\end{equation}
where $A_{\parallel} \sim 3.0$MHz and $A_{\perp} \sim 3.7$MHz. 
We assume that the states $\ket{-1,\pm \half}$ are far detuned by a dc magnetic field, 
and tune to the crossing of $\ket{0,-\half}$ and $\ket{1,\half}$, where
states are labeled by $\ket{S_z,I_z}$.
The $A_{\perp}$ term in \eqref{eq:hyperfine} 
mixes the $\ket{0,\half}$ and $\ket{1,-\half}$ states, 
yielding the energy levels 
shown versus magnetic field in Fig.~S1a. 
We now define the
states 
$\ket{A} = \beta \ket{1,-\half}  -\alpha \ket{0,\half}$,
$\ket{B} = \ket{0,-\half}$,
$\ket{C} = \ket{1,\half}$,
and $\ket{D} = \alpha \ket{1,-\half} + \beta  \ket{0,\half}$. 
To allow for resonant hops of spin excitations we work
at a point where states
$\ket{B}$ and  $\ket{C}$ are nearly degenerate,
 setting the
coefficients $\alpha = 0.531$ and $\beta = 0.847$.

\begin{figure*}
\centering
\includegraphics[width=7.00in]{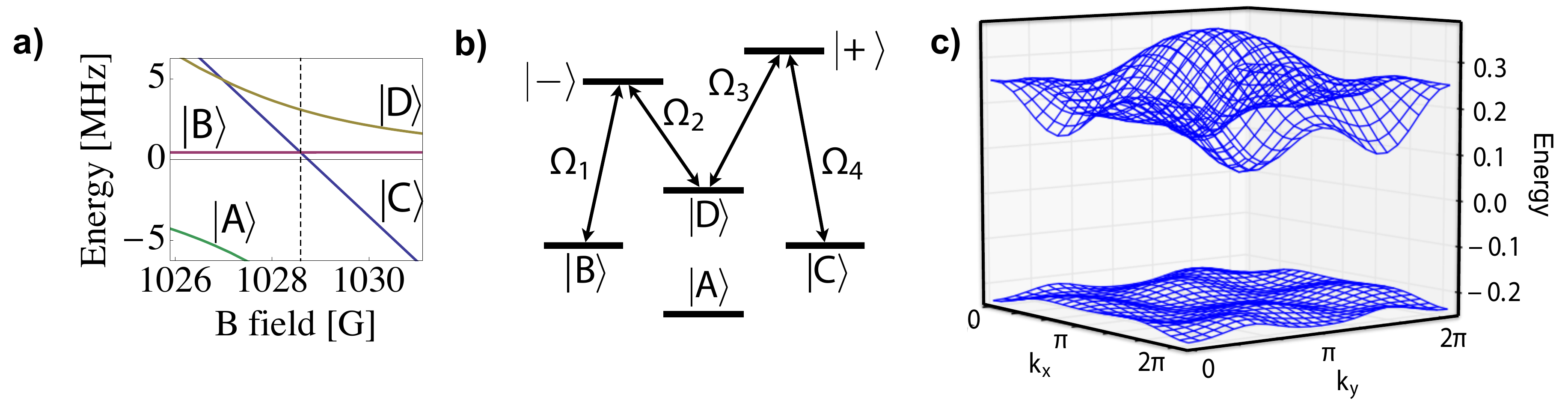}
\caption{ (a) Magnetic field required to tune the hyperfine coupled NV states to their desired resonances. (b) Optical dressing M-scheme  which enables sufficient control to realize topological flat bands as shown in (c) } 
\end{figure*}

\vspace{5mm}

\noindent The effective states we use on each NV center are $\ket{0} = | A \rangle$ and $\ket{1} = s |B\rangle + v  |C\rangle + w  |D\rangle$. The coefficients $s,v,w$ are determined via an optical ``M'' dressing scheme (Fig.~S1b) where the two
excited states are
$\ket{\pm} = \ket{E_x} \pm \ket{A_2}$, with
$\ket{E_x}, \ket{A_2}$ being two specific  electronic excited states of the NV. 
The state $\ket{1}$ is the so-called dark state of the M-scheme with $s=\Omega _2 \Omega _4/\tilde{\Omega} $, $v=\Omega _1 \Omega _3/\tilde{\Omega} $, $w=-\Omega _1\Omega _4/\tilde{\Omega} $. Note that lasers 1 and 3 must be linearly polarized,
while lasers 2 and 4 are circularly polarized. This elliptical polarization of light explicitly breaks time-reversal symmetry.

\vspace{5mm}

In the numerics presented in the main text, we use the parameterization  $s_i = \sin( \alpha_i ) \sin (\theta_i) $, $v_i = \sin( \alpha_i ) \cos (\theta_i) e^{i\phi_i} $, $w_i= \cos( \alpha_i ) e^{i\gamma_i}$ where $i \in \{a,b\}$ (recall the square lattice is partitioned into $a$ and $b$ sites). The mixing angle $\tan (\theta_i) = |s_i/v_i|$ characterizes the strength of the effective dipole moment of  $\ket{1}$, thereby determining the magnitude of the interactions. In the limit of $\theta_i \rightarrow 0$ one finds that the spin-flip excitation carries minimal weight in  $\ket{B} = \ket{0,-\half}$ and maximal weight in $\ket{C} = \ket{1,\half}$. Since the electronic spin dipole moment of $\ket{B}$ is effectively zero, this implies that the dipolar interaction strength increases as $\theta_i \rightarrow 0$. While topological flat-bands can be found for a variety of parameter regimes, we find that the clearest numerics are obtained for:
$\Theta_0 = 0.615, \Phi_0 = 5.32, \theta_a = 0.598, \theta_b =  1.051, \phi_a = 1.087, \phi_b=3.402, \alpha_a = 2.844, \alpha_b= 2.258, \gamma_a=4.089,\gamma_b =4.047$. Here the bands exhibit a flatness ratio $f \approx 8.8$ (Fig.~S1c) and phase diagrams are subsequently obtained  by varying $\Phi_0$ and $\theta_a$.

\begin{figure*}
\includegraphics[width=6.95in]{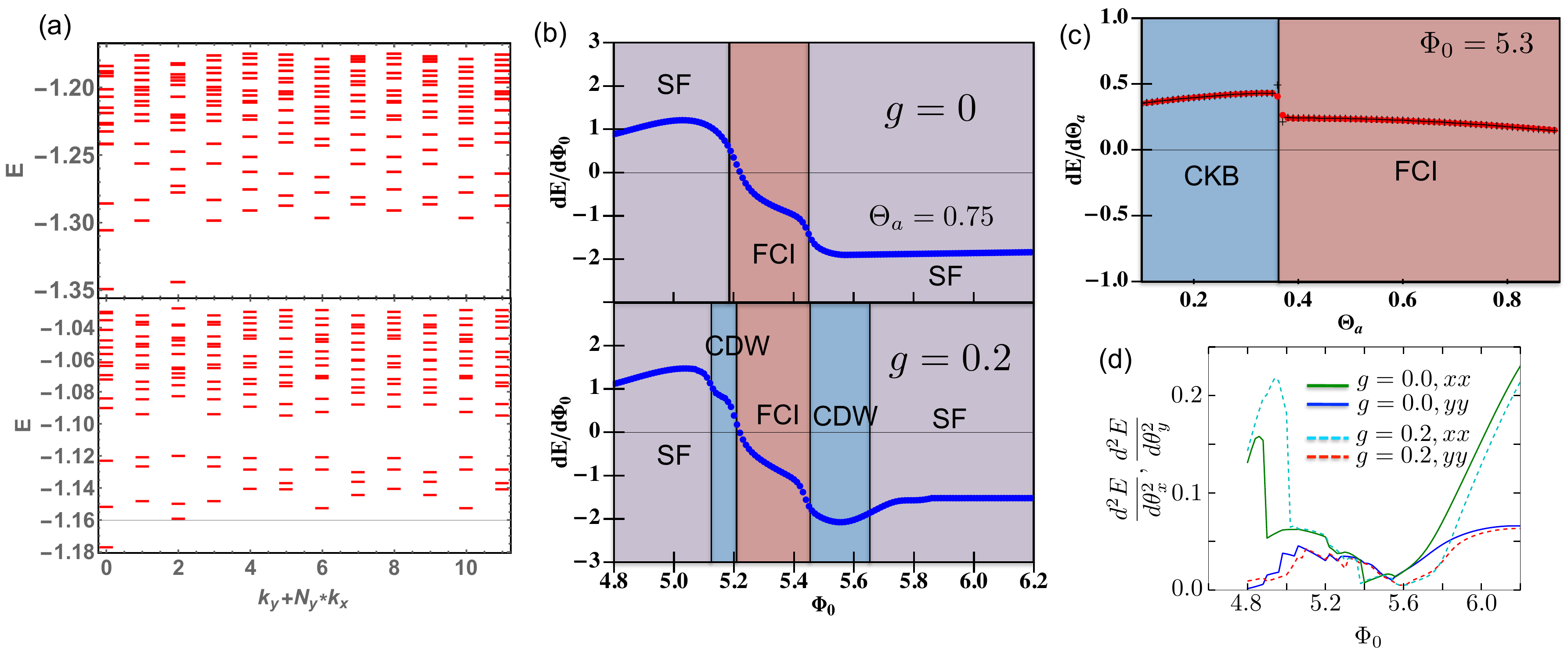}
\caption{(a) Top panel depicts the two fold ground state degeneracy in the FCI phase. Bottom panel depicts the low energy spectrum for $N_{sites}=24$, $N_{particles} = 5$. The number of total low energy states agrees with that predicted by Laughlin quasihole counting. (b) Inversion breaking response in the change in the ground state energy as a function of  $\Phi_0$ at $\theta_a=0.75$. Top panel shows $g=0$ data where one observes a SF $\leftrightarrow$ FCI  $\leftrightarrow$ SF transition. Bottom panel depicts inversion broken data, $g=0.2$, where extra kinks corresponding to the CDW occur. (c) Changes in the ground state energy as a function of   $\theta_a$  at $\Phi_0 = 5.3$ showing the CKB to FCI transition. Red circles correspond to $g=0$ and black crosses to $g=0.2$ (d) Superfluid stiffness estimated from the second derivative of the ground state energy as a function of boundary condition twists in the $\hat{X}$ and $\hat{Y}$ direction. The superfluid regions show a strong response to twists in the $x$ direction.} 
\end{figure*}

\vspace{5mm}

\noindent We now provide detailed examples of the diagnostics used to determine the many-body phases which arise at  finite lattice filling fraction. The topological features of the $\nu=1/2$ FCI require the presence of a two-fold ground state degeneracy on a torus (Fig.~S2a, top panel) as well as quasi-hole statistics which agree with a generalized Pauli principle (Fig.~S2a, bottom panel). 
As depicted in Fig.~2 of the maintext, the quantity analogous to the Hall conductance, $\sigma_{xy} =   \frac{1}{2\pi} \int \int F(\theta_x, \theta_y) d\theta_x d\theta_y =  -0.5$, appears unambiguously in the response of the system to boundary-condition twists $\{\theta_x, \theta_y\}$. 
To diagnose the CDW, we require ground state degeneracy with $\sigma_{xy} =0$. Moreover, twisting the boundary condition in either the $\hat{X}$ or $\hat{Y}$ direction does not affect the ground state energy suggesting an insulator. Finally, to diagnose a SF, we require a unique ground state. While our system sizes are too small to clearly observe the Goldstone mode of the SF, in contrast to the CDW, twisting the boundary condition dramatically alters the ground state energy (Fig.~S2d); this is consistent with a SF which harbors long range phase coherence and hence, whose energies would naturally be affected by twists in the boundary condition.

To determine rough phase boundaries between the FCI, CDW and SF, we examine the change in the ground state energy as a function of $\theta_a$ and $\Phi_0$. In particular, we expect stable phases to occur as ``smooth'' plateaus of $dE/d\theta_a$ ($dE/d\Phi_0$), while phase transitions ought manifest as jumps in $dE/d\theta_a$ ($dE/d\Phi_0$). Figure S2b,c depicts examples of ground state energy cuts in both the horizontal and vertical direction of the phase diagram. 

For a representative horizontal cut at $\theta_a = 0.75$ one indeed observes an extra kink between the FCI and SF phase upon the breaking of inversion (Fig.~S2b). In this kink region, we find a two-fold degenerate ground state in momentum sectors $(0,0)$ and $(0,\pi)$ with $\sigma_{xy}=0$. Moreover, we find that the real-space structure factor is consistent with the intervening phase being a CDW checkerboard located on the b-site sublattice. 

\section{Field theory}

Here, we provide some additional details regarding the analysis of the field theory, eq. (5) in the main text, reproduced here:
\begin{align}
\label{dual2sfcdwfqhAppendix}
\mathcal{L} &= \frac{1}{2\pi} A_e \partial a + \frac{1}{2\pi} b^\up \partial b^\down
- \frac{1}{2\pi} a \partial (b^\up + b^\down) 
\nonumber \\
&+ \sum_l [|(\partial - i b^T \tau_x) \phi_{l}|^2 - V(\{\phi_l\}),
\end{align}

As stated in the main text, this theory can simultaneously describe a superfluid, a CDW and a $1/2$ Laughlin state, depending on whether the $\phi$ fields are condensed such that they gap out both the $b$ gauge fields by the Anderson-Higgs mechanism, one of the $b$ fields or neither of the $b$ fields. 
We describe the algebraic steps leading to these identifications below.

\subsection{$1/2$ Laughlin state}

When all scalar fields $\phi_l$ are uncondensed ($\langle \phi_l \rangle = 0$) there is an energy gap to creating
excitations associated with $\phi_l$. 
Integrating them out yields only short-range interactions among the remaining fields. 
The resulting field theory is of the form
\begin{align}
\label{fqhaction1}
\mathcal{L} = \frac{1}{2\pi} A_e \partial a + \frac{1}{2\pi} b^\up \partial b^\down - \frac{1}{2\pi} a \partial (b^\up + b^\down) + \cdots,
\end{align}
where $\cdots$ include higher derivative terms for the gauge fields. 
These higher derivative terms are irrelevant compared to the Chern-Simons terms and so they can be ignored at long wavelengths.
In this limit, we may integrate out $b^\up$ to find the following constraint:
\begin{align}
\epsilon_{\mu \nu \lambda} \partial_\nu b^\down_\lambda = \epsilon_{\mu \nu\lambda} \partial_\nu a_\lambda .
\end{align}
Inserting this constraint back into (\ref{fqhaction1}) leads to
\begin{align}
\mathcal{L} = - \frac{2}{4\pi} a \partial a +  \frac{1}{2\pi} A_e \partial a  .
\end{align}
This is the well-known effective Chern-Simons field theory for the $1/2$ Laughlin state (see, e.g. \cite{wen04}). 
To verify the Hall conductance, 
we can integrate out $a$ and obtain the effective Lagrangian for the external probe field $A_e$,
\begin{align}
\mathcal{L} = \frac{1}{2} \frac{1}{4\pi} A_e \partial A_e,
\end{align}
which directly yields the $1/2$ Hall conductance,
\begin{align}
j^\mu = \frac{\delta \mathcal{L}}{\delta A_{e;\mu}} = \frac{1}{2} \frac{1}{2\pi} \epsilon^{\mu \nu \lambda} \partial_\nu A_{e;\lambda}. 
\end{align}

\subsection{Superfluid state}

Now we consider the case where both $b^\up$ and $b^\down$ are gapped by the Anderson-Higgs mechanism. 
This occurs when $\langle \phi_{l \uparrow} \rangle \neq 0 $, and $\langle \phi_{l' \downarrow} \rangle \neq 0 $,
for some $l,l'$. 
Upon integrating out $b^\uparrow$ and $b^\downarrow$, which may be accomplished at long wavelengths by simply setting $b^\alpha = 0$ in \eqref{dual2sfcdwfqhAppendix}, we obtain the effective action
\begin{align}
\mathcal{L} &= \frac{1}{2\pi} A_e \partial a - \frac{1}{g}(\epsilon_{\mu \nu \lambda} \partial_\nu a_\lambda)^2 + \cdots,
\end{align}
where we have reinstated the leading higher order Maxwell term for $a$. 
The $\cdots$ include all other terms compatible with the
gauge invariance of $A_e$ and $a$, and the lattice symmetries of the problem. 
In particular, this describes a theory of a massless 2+1 dimensional
gauge field, where fluctuations of $a$ 
physically correspond to particle density and current fluctuations, due to the coupling to the external probe field $A_e$ in the first term. 
The above theory is precisely the dual action
for a superfluid, where $a$ is dual Goldstone mode of the superfluid. 

As explained in the main text, in order to understand what additional lattice symmetries might be broken in this state, one must analyze the gauge invariant bilinears in $\phi$, the $\mathcal{O}_{k_x,k_y}^\pm$ operators, that transform non-trivially under the inversion and lattice translations.

\subsection{Insulating state}

When $\langle \phi_{l \alpha} \rangle \neq 0$ for only one choice of 
$\alpha$, then that $b^\alpha$ is gapped by the Anderson-Higgs mechanism. 
For concreteness, we consider $\alpha = \uparrow$. 
Setting $b^\uparrow = 0$ then yields the following effective 
action:
\begin{align}
%\label{dual2sfcdwfqh}
\mathcal{L} &= \frac{1}{2\pi} A_e \partial a - \frac{1}{2\pi} a \partial b^\down
%\nonumber \\
+ \sum_l [|(\partial - i b^\downarrow ) \phi_{l\downarrow}|^2 - \bar{V}(\{\phi_{l\downarrow}\}),
\end{align}
where $\bar{V}$ corresponds to the previous $V$, but with the condensed bosons replaced by $c$-numbers. 
The remaining uncondensed bosons are massive. Integrating them out yields,
\begin{align}
%\label{dual2sfcdwfqhB}
\mathcal{L} &= \frac{1}{2\pi} A_e \partial a - \frac{1}{2\pi} a \partial b^\down + \cdots
\end{align}
Now, we see that integrating out $b^\down$ will enforce a constraint at long wavelengths:
\begin{align}
\epsilon_{\mu \nu \lambda} \partial_\nu a_\lambda = 0.
\end{align}
This effectively Higgses $a$. 
Reinstating the leading higher order terms for $A_e$ gives the action
\begin{align}
\mathcal{L} = - \frac{1}{g} (\epsilon_{\mu \nu \lambda} \partial_\nu A_{e\lambda})^2 + \cdots,
\end{align}
where $\cdots$ include other terms compatible with the lattice symmetries and gauge invariance of $A_e$. 
This is the effective response theory for an insulating state, as can
be seen most simply be noting 
that the boson current $j = \frac{2}{g} \partial^2 A_e = 0$ for uniform applied fields $E = \epsilon \partial A_e$.
Moreover, all excitations of this phase are gapped, and there is no fractionalization, as expected for a topologically trivial insulator. 

Again, in order to identify the type of symmetry-breaking order in this insulator, we need to analyze the 
fate of the gauge-invariant bilinears in $\phi_l$, which transform non-trivially under the symmetries. 
From this analysis, we conclude that the insulator necessarily breaks the lattice translation symmetries
and is therefore properly identified as a CDW. 

\subsection{Broken symmetry patterns}

The above states also lead to spontaneous breaking of the lattice
symmetries. Here, we will provide some additional details of the
analysis that allow us to determine the patterns of symmetry breaking
for the superfluid and CDW state shown in Fig. 3 of the main text. A more exhaustive treatment for the full Ginzburg-Landau functional will appear in a future work. 

In order to diagnose the patterns of broken symmetry, we use the gauge-invariant bilinear operators that transform non-trivially under the lattice translational and inversion symmetries. These were described in the main text. We reproduce them here for convenience:
\begin{align}
\mathcal{O}_{0,0}^{\alpha} &\equiv \phi_\alpha^\dagger \phi_\alpha
\;\;\;
\mathcal{O}_{\pi,0}^\alpha \equiv \phi_\alpha^\dagger \sigma^z \phi_\alpha
\nonumber \\
\mathcal{O}_{0,\pi}^\alpha &\equiv \phi_\alpha^\dagger \sigma^x \phi_\alpha
\;\;\;
\mathcal{O}_{\pi,\pi}^\alpha \equiv \phi_\alpha^\dagger \sigma^y \phi_\alpha
\end{align}
The linear combination $\mathcal{O}_{k_x,k_y}^\pm \equiv \mathcal{O}_{kx,ky}^\up \pm \mathcal{O}_{k_x,k_y}^\down$ is inversion even (odd).

The two-parameter slice of the phase diagram shown in Fig. 3 used the parameters $v_2 < v_3 < 0, w_2 <0$, and $w_1, v_3, w_3 > 0$
and $u > 0$ large enough to stabilize the potential. We first consider the case where the inversion breaking parameter $g =
0$. In such a regime, one can verify that at mean-field level, when $r > 0$, all of the $\phi$ are uncondensed, all lattice symmetries are
preserved, and the system is in the FCI phase. When $r < 0$, then the system either realizes the superfluid phase (when $v_1 < v_1^c$) or
the Mott insulating CDW phase (when $v_1 > v_1^c$). At the mean field level, the critical value is $v_1^c = v_2 - 2w_1-v_3$

In the Mott insulating CDW phase, in the parameter regime described above, the minimum of the Ginzburg-Landau functional requires 
$|\phi_{0\uparrow}| = |\phi_{1\uparrow}| \neq 0, \phi_{0\downarrow} = \phi_{1\downarrow} = 0$, or vice versa 
($|\phi_{0\downarrow}| = |\phi_{1\downarrow}| \neq 0, \phi_{0\uparrow} = \phi_{1\uparrow} = 0$). Assuming the first case without loss of generality, we find that
the fact that $w_1 > 0$ further implies in this regime that $\phi_{0\uparrow} = \pm i \phi_{1\uparrow}$. Therefore, it is straightforward to verify:
\begin{align}
\langle \mathcal{O}_{0,0}^\uparrow \rangle \neq 0, &\;\;\langle \mathcal{O}_{\pi,0}^\uparrow \rangle = 0
\nonumber \\
\langle \mathcal{O}_{0,\pi}^\uparrow \rangle = 0, &\;\; \langle \mathcal{O}_{\pi,\pi}^\uparrow \rangle \neq 0,
\end{align}
while $\langle \mathcal{O}_{k_x, k_y}^\downarrow \rangle = 0$. Therefore we see that the CDW phase in this parameter regime has $(\pi,\pi)$ ordering, as observed in the numerics. 

In the superfluid phase, in the parameter regime described above, we find that the minimum of the Ginzburg-Landau functional requires 
$|\phi_{0\uparrow}| = |\phi_{1\uparrow}| = |\phi_{0\downarrow}| = |\phi_{1\downarrow}|$, and
$\phi_{0\uparrow} = \pm i \phi_{1\uparrow}$, $\phi_{0\downarrow} = \mp \phi_{1\downarrow}$. From this, we can conclude that
$\langle \mathcal{O}_{\pi,0}^\alpha \rangle = 0$, $\langle \mathcal{O}_{0,\pi}^\alpha \rangle = 0$, 
and $\langle \mathcal{O}_{\pi,\pi}^\uparrow\rangle = - \langle \mathcal{O}_{\pi,\pi}^\downarrow\rangle \neq 0$. Therefore, 
$\langle \mathcal{O}_{\pi,\pi}^+\rangle = 0 $ and $\langle \mathcal{O}_{\pi,\pi}^-\rangle \neq 0 $. This implies that the superfluid phase
has a non-zero order parameter with momentum $(\pi,\pi)$. Since this non-zero order parameter is inversion odd, it does not mix with the
density, which is inversion even. It does, however, mix with the current, which is inversion odd. We conclude that the superfluid has a
non-zero current order at $(\pi,\pi)$. Since the superfluid exists in the presence of strong time reversal symmetry breaking, it is reasonable that its ground state possesses non-zero average currents. 

When $g > 0$, it is clear that the direct FCI to SF transition will be split into two transitions, with an intervening CDW state. This is because
when $g > 0$, as we tune $r$ from positive to negative, it is more favorable to first turn on the expectation value for $\phi_{0\downarrow}, \phi_{1\downarrow}$
when $r \sim g$, and then turn on the expecation value for the remaining fields when $r \sim -g$.

\end{document}